# Anisotropic Current-Controlled Magnetization Reversal in the Ferromagnetic Semiconductor (Ga,Mn)As

Yuanyuan Li[1], Y. F. Cao[1], G. N. Wei[1], Yanyong Li[1], Y. Ji and K. Y. Wang[1,*]

[1] SKLSM, Institute of Semiconductors, CAS, P. O. Box 912, 100083, Beijing, P. R. China

K. W. Edmonds[2], R. P. Campion[2], A. W. Rushforth[2], C. T. Foxon[2], and B. L. Gallagher[2]

[2] School of Physics & Astronomy, University of Nottingham, Nottingham NG7 2RD, United Kingdom

Electrical current manipulation of magnetization switching through spin-orbital coupling in ferromagnetic semiconductor (Ga,Mn)As Hall bar devices has been investigated. The efficiency of the current-controlled magnetization switching is found to be sensitive to the orientation of the current with respect to the crystalline axes. The dependence of the spin-orbit effective magnetic field on the direction and magnitude of the current is determined from the shifts in the magnetization switching angle. We find that the strain induced effective magnetic field is about three times as large as the Rashba induced magnetic field in our GaMnAs devices.

[*] E-Mail: kywang@semi.ac.cn

The manipulation of the ferromagnetic magnetization using electrical current has attracted increased attention since it gives the opportunities to integrate magnetic functionalities into electronic circuits. Currently there are two main approaches to realize this functionality: (1) the spin polarized current induces a local torque (the spin transfer torque), which has been extensively used to study the current-driven domain wall motion and the magnetization switching of the free layer in magnetic multilayer structures [1-5]; (2) the electrical current through spin orbital (SO) interaction generates an effective magnetic field, which recently has also been used to control the magnetization and domain walls in both ferromagnetic semiconductors and ferromagnetic metals [6-10]. In the second method, it is very important to have strong SO interactions in the studied system. Rashba and Dresselhaus SO interactions are induced by structural inversion asymmetry (SIA) and bulk inversion asymmetry (BIA) respectively [11,12]. Symmetry reduction owing to uniaxial strain can also induce SO terms linear in $k$, which have the same symmetry of the linear term of the Dresselhaus SO interaction [13]. In zinc-blende III-V (100) layer semiconductors, neglecting the cubic Dresselhaus SO term, the Hamiltonian of the SO interaction can be written down as: $H_{SO} = \alpha(k_y\sigma_x - k_x\sigma_y) + \gamma(k_x\sigma_x - k_y\sigma_y)$, where α and γ denote the material dependent Rashba and Dresselhaus/strain-induced SO constants, and $\sigma_{x,y}$ are the Pauli spin matrices [14,15]. The schematic diagram of the spin–orbit induced effective magnetic field orientation with respect to the electrical current direction is shown in Fig. 1(a).

The *p-d* interactions in (Ga,Mn)As between localized Mn moments and the holes in the valence band result in large SO effects [16]. The strong SO interaction in (Ga,Mn)As not only introduces the magnetocrystalline anisotropy and the anisotropic magnetoresistance, but also can be used to control the magnetization reversal [5,6, 17-19]. It was found that the torque on (Ga,Mn)As induced by electrical currents is mainly from the strain induced SO interaction [6]. In this letter, the electrical current was applied to ferromagnetic semiconductor (Ga,Mn)As Hall bar devices along different in-plane major crystalline orientations to study the current control of the magnetization reversal. We observed that when the current density is large enough, the current induced effective magnetic field can switch the magnetization for all three device orientations investigated. However, the magnitude of the effective field induced by the electrical current depends on the direction of the electrical current applied in the studied (Ga,Mn)As devices.

The $Ga_{0.94}Mn_{0.06}As$ film, of thickness $t$=25nm, was directly grown on a semi-insulating GaAs(001) substrate by low temperature non-equilibrium molecular beam epitaxy. We then patterned 5μm wide Hall bar devices with neighbor Hall contact separation 20μm along different in-plane major crystalline axes using standard electron beam lithography and chemical wet etching. An optical image of a fabricated device is shown in Fig. 1(b). The Curie temperature was determined to be 85K by the temperature derivative of the resistivity for all these devices [20]. We measured the current induced magnetization reversal at two different situations: (1) at fixed environmental temperature T=40K; (2) at fixed device temperature $T_D$=40K, after accounting for the effect of Joule heating by the current. In the latter case, we calibrate the device temperature under an external magnetic field of 1000 Oe to minimize the effect of the anisotropic magnetoresistance in (Ga,Mn)As [19]. We obtain the device temperature $T_D$ by comparing the longitudinal resistance under large applied current to the temperature dependent longitudinal resistance measured at I = 10μA, where Joule heating is negligible. The temperature rise due to Joule heating is less than 10K for the largest applied currents.

The current induced effective field was studied by monitoring the longitudinal and transverse resistances while an external magnetic field was rotated in the plane of the film. We define $\varphi_H$ as the angle between the magnetic field and current directions. The transverse resistance can be written as $R_{xy} = (R_{\parallel} - R_{\perp})\sin 2\theta / 2$, where θ is the angle between the current and magnetization directions, and $R_{\parallel}/R_{\perp}$ is the longitudinal resistance with current parallel/perpendicular to the magnetization. In (Ga,Mn)As, the $R_{\parallel}$ is found to be smaller than $R_{\perp}$ [22]. For a rotating magnetic field of magnitude 30Oe, the magnetization angle $\theta$ does not fully track the magnetic field angle $\varphi_H$, and

$R_{xy}$ measured at 10μA applied current shows a sharp switch after $\varphi_H$ passes the [110] axis. This is due to the dominant in-plane uniaxial anisotropy, with easy axis [1-10].

Figure 2(a-c) shows $R_{xy}$ versus $\varphi_H$ for a clockwise-rotating field of 30Oe, at T=40K and with a current of $\pm 2.2\times 10^5$ Acm$^{-2}$, for the three differently oriented Hall bars. For all three devices, the large applied positive and negative currents produce a shift in the angle $\phi_H^{SW}$ where the magnetization switch occurs, indicating that the electrical current can control the magnetization reversal in all these three orientations. Both positive and negative relative shifts are observed, due to the reversal of the current induced spin-orbit effective magnetic field after 180° rotation.

We define $\Delta\varphi_H^J(J)$ as the average difference in the magnetization switching angle for positive and negative currents, for a given current density $J$. The current density dependence of $\Delta\varphi_H^J$ for the three Hall bar devices at a fixed device temperature of $T_D$=40K is shown in Fig. 3(a). The current-induced effective magnetic field can be obtained from $\Delta\varphi_H^J$ using the formula $H_{eff} = H\sin(\Delta\varphi_H^J/2)/\sin(\phi_H-\theta)$, valid for small $\Delta\varphi_H^J$ [6]. The current density dependence of $H_{eff}$ at $T_D$=40K is shown in Fig. 3(b). Both $\Delta\varphi_H^J$ and $H_{eff}$ display an approximately linear dependence on $J$, with a slope that depends on the orientation of the Hall bar. For a given $J$, the current-induced field is a factor of 2 larger for current along [110] than that for current along [1-10].

In addition to the spin-orbit coupling induced magnetic field, the applied electrical current produces an Oersted field $H_{Oe}$ which may influence the magnetization switching. The maximum value of $H_{Oe}$ is of order $Jt/2$ independent of the crystalline orientation, which is around 3 to 6 times smaller than the measured $H_{eff}$. Moreover, assuming a uniform current distribution, $H_{Oe}$ on the top and bottom surfaces of the film should be equal and opposite. Therefore, $H_{Oe}$ will assist the nucleation of magnetic domains aligned in the direction of the external magnetic field, and hence is expected to induce a small shift to lower values of $\phi_H^{SW}$ for both positive and negative electrical currents. There will also be Oersted fields in the out-of-plane direction at the edges of the structure, however this is not expected to significantly influence the switching due to the strong in-plane magnetic anisotropy.

We therefore attribute the difference $\Delta\varphi_H^J$ between magnetization switching angles for positive and negative currents to the spin-orbit induced magnetic field. The anisotropic magnitude of $\Delta\varphi_H^J(J)$ and $H_{eff}$ can be explained by the different symmetry of the strain-induced effective magnetic field and the Rashba effective magnetic field. The strain-induced and Rashba fields have the same orientation with current applied along [110] direction, perpendicular to the current direction, which favors magnetization reversal across the angle close to [110] orientation. For current along the [1$\bar{1}$0] orientation, the strain-induced and Rashba SO fields have opposite directions, so the total current-induced field is parallel or antiparallel to [110] orientation depending on their relative magnitudes. For the [100] orientation, the strain-induced field is parallel or antiparallel to [100] direction, while the Rashba field is perpendicular to the current direction. In this condition, both the Rashba and strain-induced magnetic fields partially contribute to the magnetization reversal. The large difference in $H_{eff}$ for the different orientation devices suggests that both the Rashba and strain-induced SO interactions play an important role in this system. From our experimental results, given that the effective field is of opposite sign for the [110] and [1$\bar{1}$0] oriented Hall bars, we estimate a ratio of the Rashba and strain-induced effective magnetic fields of 1:3.

In conclusion, we have investigated the spin-orbit interaction induced effective magnetic field $H_{eff}$ in Ga$_{0.94}$Mn$_{0.06}$As for electrical currents applied along three in-plane major crystalline orientations. We find that the electrical current can induce a shift of the magnetization switching angle for all three orientations. The effective current induced by the electrical current shows anisotropy: the largest effective field is obtained for current along the [110] orientation, while the lowest is for the current along the [1$\bar{1}$0] orientation. We find both Rashba and strain-induced effective magnetic fields contribute to the magnetization switching in our GaMnAs devices, with a ratio between the Rashba and strain induced effective magnetic field of about 1:3.

This work was supported by "973 Program" No. 2011CB922200, 2009CB929301, NSFC grant 11174272, 61225021 and EPSRC-NSFC joint grant 10911130232/A0402. KYW acknowledges also the support of Chinese Academy of Sciences "100 talent program".

**Figure Captions:**

Figure 1: (a) Schematic diagram to illustrate the direction of the strain induced (blue arrow) and Rashba (red arrow) spin–orbit interaction induced magnetic field for a given electrical current direction (black arrow); (b) optical image of the fabricated Hall bar device.

Figure 2: Transverse resistance $R_{xy}$ versus magnetic field angle $\phi_H$ relative to the current direction, for an external magnetic field ($H$ = 30 Oe) rotating in the plane for (a) device along $[1\bar{1}0]$ direction with current density J = ±2.2 ×10$^5$ A/cm$^2$, (b) device along [110] direction with current density J = ±2.2 ×10$^5$ A/cm$^2$, and (c) device along [100] direction with current density J = ±2.2 ×10$^5$ A/cm$^2$. Solid and open symbols are for positive and negative electrical currents, respectively.

Figure 3: The current density dependences of (a) the current-induced shift in the magnetization switching angle $\Delta\varphi_H^J$, and (b) the effective magnetic field $H_{eff}$, for devices along different orientations ($[1\bar{1}0]$ triangles, [110] diamonds, [100] circles) at $T_D$=40K. Solid and dashed lines in both (a) and (b) guide the eye.

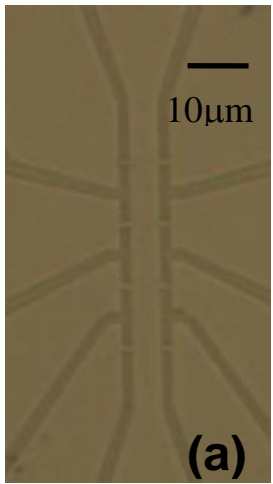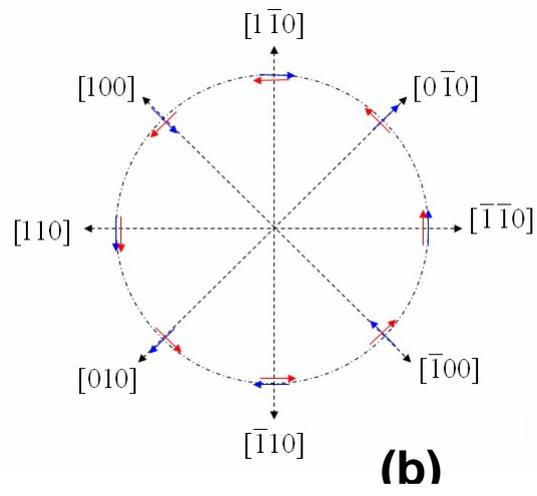

*Figure1    Yuanyuan Li et al.*

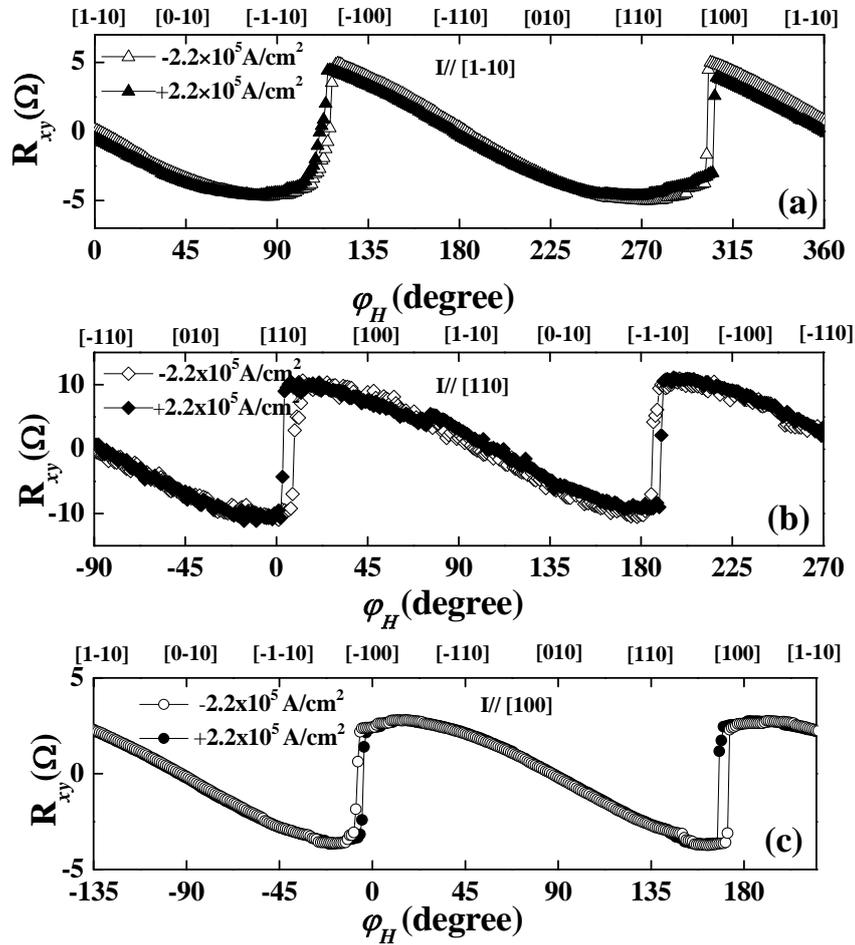

*Figure 2   Yuanyuan Li et al.*

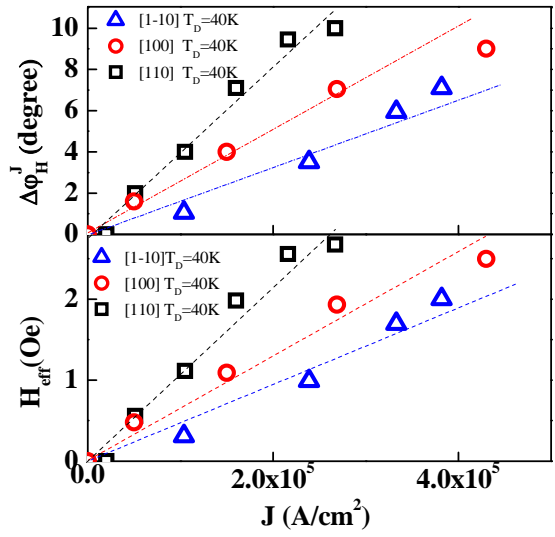

*Figure3    Yuanyuan Li et al.*